\documentclass[12pt]{article} 

\usepackage{graphicx}
\usepackage{amsmath}
\usepackage{amssymb}
\usepackage{hyperref}
\usepackage[height=8.8in,width=6.45in]{geometry}
\usepackage{tikz}
\usepackage{cite}
\usepackage{amscd}
\usepackage{setspace}
\usepackage{bbm}
\usepackage{dsfont}
\usepackage{cancel}
\usepackage{relsize}
\usepackage[utf8]{inputenc}
\usepackage{xcolor}
\newenvironment{Comment}[2]{\noindent\color{#1}{\texttt #2:}}{\par\noindent}

\numberwithin{equation}{section}

\newcommand{\PP}{{\mathbb P}}

\newcommand{\RR}{{\mathbb R}}

\newcommand{\ZZ}{{\mathbb Z}}

\newcommand{\cN}{\mathcal{N}}

\newcommand{\dd}{{\rm d}}

\newcommand{\eff}{\mathrm{eff}}

\begin{document}
\begin{titlepage}

\begin{flushright}

\end{flushright}

\vskip 3cm

\begin{center}
{\Large \bf
 On Phases of 3d ${\cal N}=2$ Chern-Simons-Matter Theories
}

\vskip 2.0cm

Wei Gu$^\infty$, Du Pei$^\infty$, Ming Zhang$^\dag$

\bigskip
\bigskip

\begin{tabular}{cc}
$^{\infty}$ Center for Mathematical Sciences and Applications, Harvard University,\\
  Cambridge, MA  02138, U.S.A.\\
$^\dag$ Department of Mathematics, the University of British Columbia, \\
 Vancouver, BC V6T 1Z2, Canada.
\end{tabular}

\vskip 1cm

 {\tt weigu@cmsa.fas.harvard.edu},
{\tt dpei@cmsa.fas.harvard.edu},
{\tt zhangming@math.ubc.ca }

\vskip 1cm

\textbf{Abstract}
\end{center}

\medskip
\noindent
We investigate phases of 3d ${\cal N}=2$ Chern-Simons-matter theories, extending to three dimensions the celebrated correspondence between 2d gauged Wess-Zumino-Witten (GWZW) models and non-linear sigma models (NLSMs) with geometric targets. We find that although the correspondence in 3d and 2d are closely related by circle compactification, an important subtlety arises in this process, changing the phase structure of the 3d theory. Namely, the effective theory obtained from the circle compactification of a phase of a 3d ${\cal N}=2$ gauge theory is, in general, different from the phase of the 3d ${\cal N}=2$ theory on $\RR^2\times S^{1}$, which means taking phases of a 3d gauge theory does not necessarily commute with compactification. We compute the Witten index of each effective theory to check this observation. Furthermore, when the matter fields have the same non-minimal charges, the 3d ${\cal N}=2$ Chern-Simons-matter theory with a proper Chern-Simons level will decompose into several identical 2d gauged linear sigma models (GLSMs) for the same target upon reduction to 2d. To illustrate this phenomenon, we investigate how vacua of the 3d gauge theory for a weighted projective space $W\mathbb{P}_{[l,\cdots,l]}$ move on the field space when we change the radius of $S^{1}$.

\bigskip
\vfill

\end{titlepage}

\setcounter{tocdepth}{2}
\tableofcontents
\section{Introduction}\label{Int}
In a groundbreaking work \cite{Witten:1993xi}, Witten provided a beautiful explanation of the connection between Verlinde algebras and the quantum cohomology of Grassmannians via the study of infrared dynamics of two-dimensional supersymmetric gauge theories.\footnote{The relation between the Verlinde algebra and the classical cohomology of the Grassmannian was first observed by Gepner \cite{Gepner:1990gr}. The refined relation was later conjectured by Vafa \cite{Vafa:1991uz} and Intriligator \cite{Intriligator:1991an}.} One effective description of the theory is as a gauged WZW model (sometimes known as $G/G$ model), whose algebra of operators realizes the Verlinde algebra (e.g.~for the group $U(N_c)$ at level $N_f-N_c$), while another is as a non-linear sigma model to a  Grassmannian (e.g.,~Gr$({N_{c}},{N_{f}}$), the moduli space of complex $N_c$ planes in $N_{f}$ space). Furthermore, the regimes where the two descriptions are individually valid can be connected by a smooth path, leading to the identification of algebras of operators in the two effective theories. This is sometimes referred to as the Verlinde/Grassmannian correspondence.

This correspondence was then better understood using 3d $\cN=2$ gauge theories by Kapustin and Willett \cite{Kapustin:2013hpk}, who point out that it is the quantum \emph{K-theory} of the Grassmannian that naturally appears in this correspondence,
\begin{equation}
    \text{Verlinde algebra of $U(N_c)_{N_f-N_c}$}\quad \simeq \quad\text{quantum K-theory of  Gr$({N_{c}},{N_{f}})$.}
\end{equation}

The goal of this paper is to refine and generalize this correspondence, and clarify an important subtlety that is needed to related the 3d and 2d perspectives. \\

The correspondence in 2d can be understood as follows. As was observed by Witten, the manifestly supersymmetric version of the gauged WZW model is
\begin{equation}\label{tes}
  S_{{\cal N}=2 }=\int d^{2}x d\theta^{+}d{\bar \theta}^{-}\widetilde{W}^{2\dd}_{\mathrm{eff}}\left(\Sigma, t\right)+c.c.,
\end{equation}
where the twisted effective superpotential $\widetilde{W}^{2\dd}_{\mathrm{eff}}\left(\Sigma, t\right)$ is a holomorphic function in terms of the twisted chiral superfield $\Sigma$ and the K\"{a}hler parameters $t$. One caveat of the supersymmetric version is that it lacks the non-abelian gauge symmetry, as $\Sigma$ field is valued in the Lie algebra of the maximal torus $T$ of the gauge group $G$. The invariance of $W^{2\dd}_{\mathrm{eff}}$ under the Weyl group $S$ of $G$ can be viewed as a remnant of the original non-abelian $G$ gauge symmetry.
However, we do not expect that it is missing any physical information because of the non-abelian/abelian correspondence both in the GWZW theory \cite{Blau:1993tv} and in the geometric phase of the supersymetric gauge theory viewed as an a gauged linear sigma models (GLSMs) \cite{Halverson:2013eua,Hori:2000kt,Gu:2018fpm,Gu:2020ana}. On the other hand, the quantum cohomology of the Grassmannian was studied by Witten in the context of the A-twisted nonlinear sigma model on Gr$(N_{c}, N_{f})$ by using tools from the GLSM. Furthermore, following \cite{Witten:1993yc,Morrison:1994fr} by summing up instanton contributions in the geometric phase of GLSM, one can argue that the effective theory (\ref{tes}) captures the full information of the quantum cohomology ring of Gr$(N_{c}, N_{f})$.\footnote{This fact has been recently re-examined in the context of supersymmetric localization of 2d GLSMs, see\cite{Closset:2015rna,Closset:2015ohf}.} As the  effective K\"{a}hler potential of the $\Sigma$ field in the geometric phase is $Q_{A}$-exact,
it is not surprising to see that Witten's analysis can be understood by studying a single effective theory (\ref{tes}) defined on the whole K\"{a}hler moduli. The Verlinde formula and the quantum cohomology of the Grassmannian are defined on the K\"{a}hler moduli near  $t=-\infty$ and $t=\infty$ respectively.

The gauged WZW model is equivalent to a pure Chern-Simons theory compactified on $S^{1}$, with Wilson loops wrapped on $S^{1}$ give rise to the local operator in the $G/G$ model. Thus there can be a  three-dimensional lift of Witten's analysis. Indeed, it was realized by Kapustin and Willett in \cite{Kapustin:2013hpk} with the starting point being a ``big theory":\footnote{In this paper we will also use the terminology ``3d gauged linear sigma model" (``GLSMs'') interchangeably with ``3d ${\cal N}=2$ Chern-Simons-matter theory,'' especially when we want to emphasize the target space perspective.}
\begin{align}\label{bth}
  3\dd\; &{\cal N}=2\ U(N_{c})\  \text{super-Chern-Simons\ at  level\ $-\frac{N_{f}}{2}$}   \\\nonumber
  + &\text{ $N_{f}$\  fundamental   chiral multiplets \  $\Phi_{i=1,\ldots,N_{f}}$}.
\end{align}
 The vacuum structure of this theory depends on the FI parameter $\zeta$ associated with the $U(1)$ subgroup of the  $U(N_{c})$ gauge group. More specifically, if $\zeta>0$, the vacuum is the Grassmannian. Expanding fields of the ``big theory" around the vacuum and integrating out massive modes, one can then obtain an effective theory: 3d non-linear sigma model (NLSM) on the Grassmannian. If $\zeta<0$, the effective theory would be the 3d ${\cal N}=2\ U(N_{c})$ pure Chern-Simons theory with level $-N_{f}$. Then, after compactifying  these two different effective theories on $S^{1}$, one expects the 3d NLSM on the Grassmannian to give the quantum K-theory of the Grassmannian,\footnote{There are different versions of quantum K-theory in the mathematical literature, and the ``right'' version here would be the quantum K-theory with a ``level structure" in the sense of \cite{Zhang1:2018}.} therefore relating it to the Verlinde algebra realized as the algebra of line operators on the 3d pure Chern-Simons theory side. Then  Witten's beautiful 2d analysis  can be recovered by taking the vanishing limit of the radius of $S^{1}$.

 Another way to see the connection is that we can first compactify the ``big theory" on $S^{1}$, and then integrate out matter fields, leading to
\begin{equation}\label{tes2}
  S_{{\cal N}=2 }=\int d^{3}x d\theta^{+}d{\bar \theta}^{-}\widetilde{W}^{3\dd}_{\mathrm{eff}}\left(\Sigma, t\right)+c.c.
\end{equation}
The three-dimensional twisted effective superpotential can be treated as the 3d lift of the two-dimensional one in (\ref{tes}). Just as the 2d case, Witten's analysis can be understood with a single three-dimensional effective theory.\footnote{A related development along these lines is the Bethe/Gauge correspondence proposed by Nekrasov and Shatashvili in \cite{Nekrasov:2009uh}, and the relation between integrable system and GWZW can be found in \cite{Gorsky:1993dq}. The problem of lifting 2d effective theories to 3d has recently attracted considerable interest due to the development of the 3d-3d correspondence (see e.g.~\cite{ Dimofte:2010tz,Dimofte:2011ju,Chung:2014qpa}). See also \cite{Cecotti:2013mba} for the study of 3d $tt^{\ast}$ geometry.
}

One natural extension of the correspondence discovered by Witten and its 3d lift by Kapustin and Willett involves complexifying $G$ to $G_{\mathbb{C}}$ on the Verlinde algebra side. Then the resulting ``equivariant Verlinde algebra" is predicted to be identified with the equivariant quantum K-theory of vortex moduli space, both realized as phases of 3d ${\cal N}$=2 Chern-Simons matter theory with a massive adjoint multiplet \cite{Gukov:2015sna}. While another natural
extension in the two-dimensional situation is to consider the superpotential in the GLSM. If the contribution from the superpotential to vacua at $\Sigma=0$ is trivial, i.e., it has at most one vacuum, then Witten's analysis still holds. More specifically, if there is no SUSY vacuum at $\Sigma=0$, the effective theory (\ref{tes}) is already enough;
while if there is one vacuum configuration at $\Sigma=0$, we can simply include this solution into the vacuum equation:
\begin{equation}\label{tesv}
  \exp\left(\Sigma\cdot\frac{\partial \widetilde{W}_{\eff}}{\partial\Sigma}\right)=1.
\end{equation}
This idea was used in \cite{Gu:2020oeb,Gu:2020zpg} to compute the quantum cohomology of Lagrangian Grassmannians, matching the results obtained from a different approach in the mathematical literature. And if one can lift the two-dimensional effective theory to a 3d theory on $\RR^2\times S^{1}$ with the effective theory (\ref{tes2}), one could then obtain the quantum K-theory of Lagrangian Grassmannians, which would produce new mathematical results. It comes to a natural question: do we have an intrinsic three-dimensional quantum field theoretic understanding of this extension?

This main goal of this paper is to investigate the above question, and the summary of our observations are as follows
\begin{itemize}
    \item \textit{Effective theories obtained from circle reductions of phases of a 3d $\mathcal{N}=2$ gauge theories on $S^{1}$ are, in general, different from phases of the 3d theory on $\RR^2\times S^{1}$.\footnote{The authors in \cite{Aharony:2013dha} studied the compactification of 4d supersymmetric gauge theories on $\RR^3$ times a finite circle. In their context, taking the effective theory of a gauge theory does commute with the compactification.}}
    \item \textit{A 3d abelian Chern-Simons-matter field theory with the same non-minimal charges $l$ and a proper Chern-Simons level will decompose into $l$ identical copies of 2d GLSMs for the same target in the vanishing radius limit of the KK reduction.\footnote{Related phenomenons have been studied in the previous literature. The dynamical decomposition of a 3d gauge theory into several different 2d gauge theories in the vanishing radius limit of $S^{1}$ \cite{Aharony:2017adm}, and see \cite{Pantev:2005zs, Hellerman:2006zs} for the investigation on the decomposition of a 2d nonlinear sigma model on a weighted projective space $W\mathbb{P}_{[l,\cdots,l]}$ into $l$ identical 2d nonlinear sigma models on the same projective space.}}
\end{itemize}

Although to better understand and generalize the Verlinde/Grassmannian correspondence is a major motivation that leads to this work, the above striking observations apply more broadly.

\subsection{Outline}
 The rest of this paper is organized as follows. In section \ref{PR}, we review some basics of 3d/2d ${\cal N}=2$ GLSMs following \cite{Intriligator:2013lca}, and we mainly focus on how the original analysis by Witten can be understood using gauge theory. In section \ref{P3GLSM}, we give a systematic study of phase structures of the three-dimensional GLSM with superpotential. Our three-dimensional GLSM is a UV-fundamental theory like we studied in the two-dimensional situation \cite{Witten:1993yc}. However, unlike the 2d case, in 3d we have a perturbative renormalizability constraint for the Lagrangian in the gauge theory. More specifically, we restrict to superpotentials that are polynomials of fields of degree at most four. We compute the Witten index of effective theories of different phases of GLSMs to justify our observations mentioned in the introduction. Furthermore, a detailed computation of the Witten index of 2d/3d Landau-Ginzburg orbifolds can be found in appendix \ref{TWLGO}. In section \ref{DKK}, we discuss the situation that the 3d GLSM for a geometric target space decomposes into several identical 2d theories for the same target space in the 2d limit. The last section is devoted to conclusions and future directions.

\section{Preliminaries} \label{PR}
In this section, we review some basic facts about 3d $\mathcal{N}=2$ supersymmetric gauge theories. We start with 3d $U(1)$ abelian gauge theories following \cite{Intriligator:2013lca}, which can be readily extended to more general cases with the geometric target being toric varieties \cite{Dorey:1999rb, Aganagic:2001uw}.
\subsection{3d ${\cal N}=2$ $U(1)$ GLSMs}\label{AbGLSM}
Consider a $U(1)$ gauge theory with classical Chern-Simons term at level $k$,\footnote{One can fix the level by requiring the A-twisted NLSM to give Givental-Lee's quantum K-theory \cite{Givental:2000,Lee:2004} of the geometric target.} Fayet-Iliopoulos (FI) term $\zeta$, and matter fields $\phi_{i}$ of charge $Q_{i}$ having no common integer factor, and real mass $m_{i}$. The consistency condition from compact group $U(1)$ implies
\begin{equation}\label{CSL}
  k+\frac{1}{2}\sum_{i}Q^{2}_{i}\in\mathbb{Z}.
\end{equation}
For the special case of one matter field with charge 1, the above condition becomes
\begin{equation}\label{}
  k=-\frac{1}{2}+\overline{k},\qquad \overline{k}\in\mathbb{Z},
\end{equation}
where the scheme ambiguity $\overline{k}$ is due to the unphysical ``Chern-Simons contact term".\footnote{The integer level Chern-Simons term can be regarded as a local contact term \cite{Closset:2012vp}.} For a generic $\zeta$ and $m_{i}$, the unbroken global symmetry is $U(1)_{R}\times U(1)_{T}\times \prod_{i}U(1)_{i}/U(1)$. Here $U(1)_{R}$ is the R-symmetry while $U(1)_{T}$ is the topological symmetry associated with the $U(1)$ gauge symmetry with the topological current given by $J_{T}=\frac{1}{4\pi}\epsilon^{\mu\nu\rho}F_{\mu\nu}$.  The parameters $m_{i}$ and $\zeta$ can be identified with background values of scalars in $\mathcal{N}=2$ vector multiplets containing classical gauge fields for these global symmetries. They are renormalized at most at one loop.

The system has richer phase structures by turning on real masses. With real masses, for example, it could have a compact Coulomb branch. Furthermore, the spectrum of non-perturbative degrees of freedom depend on real masses, which is important for understanding 3d mirror symmetry \cite{Intriligator:1996ex,Kapustin:1999ha, Aganagic:2001uw}. However, in this paper, we are interested in phases of gauge theories with a  superpotential, which means real masses are zero.

The semi-classical effective potential of the theory is
\begin{equation}\label{SCEP}
  U_{s.c.}=\frac{e^{2}_{\eff}}{2}\left(\sum_{i}Q_{i}|\phi_{i}|^{2}-\zeta-k_{\eff}\sigma\right)^{2}+\sum_{i}\left(Q_{i}\sigma\right)^{2}|\phi_{i}|^{2}.
\end{equation}
We have omitted the fermions in the above expression, as their vacuum expectation values must be zero for the spacetime Lorentz symmetry to be unbroken. The quantum-corrected parameter $k_{\eff}$ is given by
\begin{equation}\label{QCPK}
  k_{\eff} =  k+\frac{1}{2}\sum_{i}Q^{2}_{i}{\rm sign}\left(Q_{i}\sigma\right),
  \end{equation}
  and define
  \begin{equation}
    F(\sigma):=\zeta+k_{\eff}\sigma= \zeta+  k\sigma+\frac{1}{2}\sum_{i}Q_{i}|Q_{i}\sigma|.
\end{equation}
 If one turns on real masses, the FI parameter $\zeta$ also receives quantum correction.

 The vacuum configuration can be found by solving $U_{s.c.}=0$, which gives
 \begin{equation}\label{3svc}
   \sum_{i}Q_{i}|\phi_{i}|^{2}=F(\sigma),\qquad {\rm and} \qquad \left(Q_{i}\sigma\right) \phi_{i}=0, \quad {\rm for\ all}\ i.
 \end{equation}
Then, there are three different types of vacua:
\begin{itemize}
    \item Higgs phase: $\sum_{i}Q_{i}\langle|\phi_{i}|^{2}\rangle=\zeta$, and $\langle\sigma\rangle=0$. If $Q_{i}\zeta>0$ for all $i$, the vacuum configuration is a compact geometry, while if $Q_{i}\zeta<0$ for some of $j$, one has a non-compact Higgs branch.
    \item Coulomb branch: continuous vacua with $\langle\phi_{i}\rangle=0$ for all $i$, and $F(\sigma)=0$. The existence of flat direction parametrized by the $\sigma$-field requires $\zeta=k_{\eff}=0$. If one turns on real masses, it could have a continuous compact Coulomb branch. Furthermore, one can dualize the gauge field into a chiral superfield, which has a charge $q_{T}=1$ under the topological symmetry $U(1)_{T}$. Therefore, the unbroken $U(1)_{\text{gauge}}$ is equivalent to a spontaneously broken $U(1)_{T}$ global symmetry. Finally, we want to briefly mention that this phase is usually referred to as a conifold point on the K\"{a}hler moduli of a 2d GLSM where correlation functions diverge.
    \item  Topological vacua: when $F(\sigma_{I})=0$ and $k_{\eff}\neq 0$. There are isolated massive vacua at $\langle\phi_{i}\rangle=0$ and $\sigma_{I}=-\zeta/k_{\eff}$. The matter fields are all heavy and integrated out, then it reduces to a low energy effective Chern-Simons theory $U(1)_{k_{\eff}}$.
\end{itemize}

A more detailed discussion on phase structures with real masses can be found in \cite{Intriligator:2013lca}. In the next section, we discuss 3d non-abelian gauge theories and mainly focus on the extension of Witten's analysis into three dimensions first pioneered by Kapustin and Willett \cite{Kapustin:2013hpk}.

\subsection{3d ${\cal N}=2$ $U(k)$ GLSMs}\label{GRGLSM}
We start with the ``big theory" mentioned in the introduction: 3d ${\cal N}=2$ $U(N_{c})$ gauge theory with $N_{f}>N_{c}$ fundamental chiral superfields $\Phi_{i}$, $i=1,\ldots,N_{f}$, the action of the vector multiplet is the ${\cal N}=2$ Chern-Simons theory with level $-\frac{N_{f}}{2}$. Then, the effective theories can be in two different phases, depending on the sign of the now real-valued FI parameter $\zeta$:
\begin{itemize}
\item If $\zeta>0$, the vacuum configurations can be found by solving
\begin{equation}\label{Grv}
  \sum^{N_{f}}_{i=1}\left\langle\overline{\phi}^{ia}\phi_{ib}\right\rangle=\zeta\delta^{a}_{b},\qquad {\rm and}\quad \langle\sigma\rangle=0.
\end{equation}
As expected, the vacua are parametrized by the Grassmannian Gr$(N_{c}, N_{f})$. Then integrating out massive modes leads to a 3d NLSM onto the Grassmannian in the infrared.

\item If $\zeta<0$, we have topological vacua. More specifically, solving
\begin{equation}\label{Grv2}
  F(\sigma)=\zeta-\frac{N_{c}}{2}(\sigma_{i}-|\sigma_{i}|)=0, \quad i=1,\ldots,N_{c},\qquad {\rm and} \quad \langle\phi_{ib}\rangle=0
\end{equation}
gives a unique solution $\sigma_{i}=-\frac{\zeta}{N_{c}}>0$ for all $i$. The low-energy effective theory is then the 3d ${\cal N}=2$ Chern-Simons theory with level $-N_{f}$.
\end{itemize}
This is how Kapustin and Willett explained Witten's correspondence using gauge theories in three dimensions.

Here we have omitted the discussion of the non-compact Coulomb branch located at $\zeta=0$ and this could lead to one important caveat: as $\zeta=0$ is a real codimension-1 singularity of the parameter space, the theory may undergo a genuine phase transition when the sign of the FI parameter $\zeta$ flips. The two phases are not obviously connected by a smooth path in a bigger parameter space of the 3d theory, which means that even the number of vacua or Witten index in principle can change when the theory undergoes phases transition. For example, there could be additional vacua coming from infinity or disappearing to infinity in the field space, further complicating the 3d version of the correspondence in 2d discovered by Witten. Fortunately, it is not the case for the examples we discussed in this paper. One can argue that for quantities that can be defined using the effective 2d theory on $\RR^2\times S^1$, for which the FI parameter complexifies and a continuous path connecting the two phases appears, they have to be the same before and after the phase transition. For example, the Witten index of the 3d $\mathcal{N}=2$ theory can be defined as a partition function on a three-torus, and can be regarded as that of the 2d ${\cal N}=(2,2)$ effective theory on $T^2$, and is thus the same for the two different phases \cite{Witten:1993yc}.\footnote{In fact, the invariance of Witten index under circle compactification was already used by Witten in \cite{Witten:1982df} to reduce the theory to the computation of vacua of a 0+1 dimensional SUSY theory on target spaces.}

To better scrutinize the process of $S^1$ compactification, we next review some aspects of 3d gauge theories on $\RR^2 \times S^{1}$.

\subsection{3d $\cN=2$ gauge theories on $\RR^2\times S^{1}$} \label{GLSM32}
 Assuming the circle $S^1$ has radius $R$, it gives rise to a 2d ${\cal N}=(2,2)$ gauge theory with an infinite tower of massive matter fields. The physical 2d FI parameter depends on the worldsheet\footnote{In this paper, we are only concerned with quantum field theories decoupled from gravity. However, we will still sometimes employ the terminology ``worldsheet'' to refer to the spacetime of a 2d QFT as it is widely used.} physical scale $\mu$ and the dynamical scale $\Lambda\sim\exp\left(-2\pi R\zeta\right)/R$
\begin{equation}\label{}\nonumber
  r=\sum_{i} \rho_{i}\log\left(\mu/\Lambda\right)=2\pi R\zeta+\sum_{i} \rho_{i}\log\left(\mu R\right),
\end{equation}
where $\rho_{i}$ denote gauge charges of matter. Now consider the worldsheet IR $\mu\ll \Lambda$ i.e. $r\ll 0$. Then the vacuum expectation values of $\sigma$-fields are large, and $\sigma_{a}\sim\sigma_{a}+\frac{i}{R}$ due to large gauge transformations along $S^{1}$. This means matter fields are heavy. Once they are integrated out, the theory reduces to an effective theory defined by the effective twisted superpotential
\begin{eqnarray}\label{3tefs}
  \widetilde{W}^{3\dd}_{\eff}\left(\Sigma_{a}, t_{a}\right)\cdot 2\pi R &=&\frac{1}{2}k^{ab}\left(\ln x_{a}\right)\left(\ln x_{b}\right)+\sum_{a}\left(\ln q_{a}\right)\left(\ln x_{a}\right) \\\nonumber
 &&+ \sum_{a}\left(i\pi\sum_{\alpha>0}\alpha^{a}_{\mu}\right)\left(\ln x_{a}\right)+\sum_{i}\left[{\rm Li}_{2}\left(x^{\rho_{i}}\right)+\frac{1}{4}\left(\rho_{i}\ln x\right)^{2}\right].
\end{eqnarray}
In the above expression, $i$ indexes fields, $x_{a}=\exp\left(-2\pi R\Sigma_{a}\right)$, and $x^{\rho_{i}}=\prod_{a}x_{a}^{\rho^{a}_{i}}=\exp\left(-2\pi  R\rho^{a}_{i}\Sigma_{a}\right)$. Because of their three-dimensional origins, the $\Sigma_{a}$'s are periodic, with periodicity which we will take to be
\begin{equation}\label{}\nonumber
  \Sigma\equiv\Sigma+\frac{i}{R}.
\end{equation}
For the expression with non-zero real masses turned on, see \cite{Nekrasov:2009uh,Ueda:2019qhg}. Let us apply the formula (\ref{3tefs}) to theories reviewed in section \ref{AbGLSM} and \ref{GRGLSM}.

For gauge theories in section \ref{AbGLSM}, we have $\rho_{i}=Q_{i}$, and no roots $\{\alpha\}=\emptyset$
\begin{equation}\label{AB3tesfs}
   \widetilde{W}^{3\dd}_{\eff}\left(\Sigma, t\right)\cdot 2\pi R=\frac{1}{2}\left(k+\frac{\sum_{i}Q_{i}Q_{i}}{2}\right)\left(\ln x\right)^{2}+\left(\ln q\right)\left(\ln x\right)+\sum_{i}{\rm Li}_{2}\left(x^{Q_{i}}\right).
\end{equation}
One can find the effective Chern-Simons level is
\begin{equation}\label{csle}
  \widetilde{k}_{\eff}= k+\frac{\sum_{i}Q_{i}Q_{i}}{2}.
\end{equation}
By comparing \eqref{csle} to  (\ref{QCPK}), we find that the quantum correction for parameters in gauge theories on $\RR^2\times S^{1}$ is different from the one on flat spacetime. The sign of effective real masses from the $\sigma$ field is not important in gauge theories on $\RR^2\times S^{1}$. Furthermore, one can find that equation (\ref{AB3tesfs}) always have solutions with different signs corresponding to effective real masses.

The equation (\ref{csle}) was first used in \cite{Aganagic:2001uw} to understand 2d mirror symmetry from 3d mirrors, and in that paper they chose
\begin{equation}\label{}\nonumber
  k=-\frac{\sum_{i}Q_{i}Q_{i}}{2}
\end{equation}
where $\widetilde{k}_{\eff}$ vanishes. This choice gives Givental-Lee's quantum K-theory \cite{Givental:2000,Lee:2004} of a toric variety. Briefly, quantum K-theory arises as the OPE algebra of Wilson loops about the $S^{1}$ under the partial topological twisted along $\RR^2$ \cite{Aganagic:2017tvx}.

For the case of the gauge group $U(N_{c})$, it is easy to compute that
\begin{equation}\label{}\nonumber
   \sum_{\alpha>0}\alpha^{a}_{\mu}\sim N_{c}-1
\end{equation}
for all $a$, so it changes $q$ by a phase factor $(-)^{N_{c}-1}$. Notice that we can divide the Chern-Simons level $k_{U(N_{c})}$ into $k_{U(1)}$ and $k_{SU(N_{c})}$. Following \cite{Gu:2020zpg}, we write the twisted effective superpotential as
\begin{eqnarray}\label{GRTE}\nonumber
  \widetilde{W}^{3\dd}_{\eff}\left(\Sigma , t \right) \cdot 2\pi R &=& \frac{1}{2}k_{SU(N_{c})}\sum_{a}\left(\ln x_{a}\right)^{2}+\frac{k_{U(1)}-k_{SU(N_{c})}}{2N_{c}}\left(\sum_{a}\ln x_{a}\right)^{2}\\
  &&\hspace{-0.5cm}+\left(\ln(-)^{N_{c}-1}q\right)\sum_{a}\ln x_{a}+N_{f}\sum_{a}{\rm Li}_{2}\left(x_{a}\right)+\frac{N_{f}}{4}\sum_{a}\left(\ln x_{a}\right)^{2}.
\end{eqnarray}
To give the quantum K-theory of a Grassmannian \cite{Buch:2008,Jockers:2019lwe,Ueda:2019qhg,Gu:2020zpg}, the levels are chosen to be:
\begin{equation}\label{GRLV}
  k_{U(1)}=-\frac{N_{f}}{2},\qquad k_{SU(N_{c})}=N_{c}-\frac{N_{f}}{2}.
\end{equation}
This choice of levels can be understood from the point of view of mass deformation of a 3d ${\cal N}=4$ gauge theory, or geometrically, localizing onto K\"ahler fixed loci of an isometry acting on a hyper-K\"ahler space (cf.~\cite{Bullimore:2019qnt}). The term $ N_{c}$ arises from integrating out the extra ${\cal N}=2$ chiral multiplet needed to build the ${\cal N}=4$ vector multiplet, and the $-N_{f}/2$ from integrating out half of the ${\cal N}=4$ hypermultiplets. More details about how to get quantum K-theory from formula (\ref{GRTE}) with these levels (\ref{GRLV}) can be found in \cite{Ueda:2019qhg,Gu:2020zpg}.

One may notice that the levels~(\ref{GRLV}) corresponding to Givental-Lee's quantum K-theory of a Grassmannian are different from those proposed by Kapustin and Willett. In fact, it was first observed by Ruan and the last author in \cite{Zhang:2018} that the Verlinde formula corresponds to quantum K-theory with a ``level structure".\footnote{It was already understood in  \cite{Witten:1993xi} that correlation functions of gauged WZW models can be mapped to correlation functions of the Gr$(k, N)$ by inserting an operator $\left(\det \sigma\right)^{-(g-1)(N-k)}$ in the path integral of NLSM on Gr$(k, N)$. This extra factor $\left(\det \sigma\right)^{k}$ is a remnant of the ``level structure" defined in the context of \cite{Zhang1:2018}.
} The mismatch between the Chern-Simons level $k_{SU(N_{c})}$ for Givental-Lee's quantum K-theory and the one proposed by Kapustin and Willett agrees with that observation.

So far we have reviewed some well-understood results in the literature. We now present a simple extension of Witten correspondence based on the physics of 2d ${\cal N}=(2,2)$ in the next section.

\subsection{A simple extension of Witten's correspondence}\label{SEW}
 Let us start with a concrete example. Consider a 3d GLSM for a degree-2 hypersurface in a projective space $\mathbb{P}^{N}$ for $N>3$ denoted by $\mathbb{P}^{N}[2]$, with the Chern-Simons level $k$ chosen to be $-\frac{N-3}{2}$.  Let us compactify the gauge theory on $S^{1}$ first. The geometric phase, at $r\gg0$, is a 3d NLSM on $\mathbb{P}^{N}[2]$, whose Witten index of NLSM on $\mathbb{P}^{N}[2]$, is $N$ if $N$ is even, and $N+1$ if $N$ is odd. The ``LG point''
 at $r\ll0$ has two sectors: the topological vacua, at $\sigma\neq0$, are given by the critical points of the  twisted effective superpotential
 \begin{equation}\label{AB2PT}
   \widetilde{W}_{\eff}\left(\Sigma, t\right)\cdot 2\pi R=\frac{1}{2}\left(k+\frac{N-3}{2}\right)\left(\ln x\right)^{2}+\left(\ln q\right)\left(\ln x\right)+\left(N+1\right){\rm Li}_{2}\left(x \right)+{\rm Li}_{2}\left(x^{-2} \right).
 \end{equation}
 The critical locus equation, defining the vacua, is
 \begin{equation}\label{VEAB2P}
   \left(1-x\right)^{N-1}=q\left(1+x\right)^{2},
 \end{equation}
 which has $N-1$ solutions. The other sector is, at $\sigma=0$, up to KK modes, essentially a 2d $\mathbb{Z}_{2}$-orbifolded Landau-Ginzburg (LG) model:
\begin{equation}\label{LG2}
   W=\sum^{N+1}_{i=1}\Phi^{2}_{i}.
 \end{equation}
The ground states can be computed from the Hamiltonian:
\begin{equation}\label{GS2LG}
 |\Omega\rangle,\qquad |{\rm TS}\rangle.
\end{equation}
Here the twisted-sector $|{\rm TS}\rangle$ is $\mathbb{Z}_{2}$ invariant, while the untwisted-sector, represented as the top form $ d\phi_{1}\wedge\cdots\wedge d\phi_{N+1}$, transforms under the $\mathbb{Z}_{2}$ action as follows
\begin{equation}\label{}\nonumber
  |\Omega\rangle\mapsto (-)^{N+1}|\Omega\rangle.
\end{equation}
So only for odd $N$, we have a gauge-invariant untwisted ground state. The Witten index at the LG point is the sum of the ground states of the two sectors, and it is the same as the Witten index in the geometric phase. One then readily observes that, as mentioned in the introduction, for even $N$, the twisted effective superpotential captures the whole quantum K-theory of the geometric target by a simple modification of the vacuum equation (\ref{VEAB2P}) as follows
\begin{equation}\label{}\nonumber
  \left(1-x\right)^{N-1}\left(1-x\right)=q\left(1+x\right)^{2}\left(1-x\right).
\end{equation}
This can also happen in non-abelian GLSMs, which has been used in \cite{Gu:2020oeb,Gu:2020zpg} for computing the quantum cohomology and K-theory of the Lagrangian Grassmannian. Thus, for a GLSM on $\RR^2\times S^{1}$, an extension of Witten's analysis still holds.

One may naturally expect that the extended Witten's story was still true in the three-dimensional flat spacetime. However, it is striking that the flat spacetime physics of gauge theory with superpotential is different from what we expected from the intuition of 2d gauge theory. To make our observations more clear, we will give a detailed study of phases of the three-dimensional GLSM with superpotential in the next section.

\section{Phases of 3d ${\cal N}$=2 Chern-Simons-matter theories with superpotentials}\label{P3GLSM}

Before we give a detailed study of 3d supersymmetric gauge theories, let us start with a 3d Landau-Ginzburg model (or often referred to as the Wess-Zumino model) first to see the constraint on the superpotential \cite{Aharony:1997bx} from 3d perturbative quantum field theory. The Lagrangian of a 3d LG model is
\begin{equation}\label{3dWZ}
  \int d^{4}\theta K\left(\overline{\Phi},\Phi\right)+\left(\int d^{2}\theta W\left(\Phi\right)+c.c\right).
\end{equation}
One can read off that the mass dimension of the chiral superfield $\Phi$ is 1/2 from the K\"{a}hler potential. So unlike the case in 2d, where one can study any holomorphic superpotential $W$, in the three-dimensional case the marginal superpotential is $\Phi^{4}$. The orbifolded LG model in three-dimension can also be studied,  
but it seems that one can not give a definition of the three-dimensional GLSM for Calabi-Yau 3-fold targets, such as the quintic threefold. However, we make two comments here. First, in practice, there are still many interesting Calabi-Yau targets that can be constructed with a gauge theory Lagrangian. Second, one might still study 3d sigma models on quintic and other examples which do not admit UV gauge theory descriptions. We focus on the first point in this paper and leave the second to future work.

\subsection{The abelian case}\label{ABS}
Consider a $U(1)$ gauge theory with $N+1$ charge-one fields $\Phi_{i}$, $i=1,\ldots,N+1$, and one $P$-field with a negative charge $-d$. We choose the bare Chern-Simons level to be $k=-\frac{N+1-d^{2}}{2}$, and a superpotential
\begin{equation}\label{}\nonumber
  W=PG\left(\Phi_{i}\right).
\end{equation}
Because of the $P$-field in the superpotential, we impose the following condition on the degree:
\begin{equation}\label{}\nonumber
  1\leq d\leq3.
\end{equation}
 The special choice of the Chern-Simons level guarantees that the topological vacua will appear in only one phase. One can, in principle, choose any bare Chern-Simons level as long as it is consistent with gauge invariance. However, the analysis with a more general choice of the level will be completely analogous to the one presented below, and does not require any additional new ingredients.

The semi-classical effective potential of the theory is
\begin{eqnarray}\label{SCEP2}\nonumber
  U_{s.c}&=&\frac{e^{2}_{\eff}}{2}\left(\sum^{N+1}_{i=1}|\phi_{i}|^{2}-d|p|^{2}-\zeta-k_{\eff}\sigma\right)^{2}+\sum^{N+1}_{i=1}\sigma^{2}|\phi_{i}|^{2}+d^{2}\sigma^{2}|p|^{2}\\
  &&+\left|G\left(\phi_{i}\right)\right|^{2}+\sum^{N+1}_{i=1}\left|p\frac{\partial G\left(\phi_{i}\right)}{\partial\phi_{i}}\right|^{2}.
\end{eqnarray}

The structure of the vacua is as follows:

\begin{itemize}
\item If $\zeta>0$, we have a smooth geometric vacuum configuration
\begin{equation}\label{GAGLSM}
  \sum^{N+1}_{i=1}\left\langle|\phi_{i}|^{2}\right\rangle=\zeta,\qquad \left\langle \sigma\right\rangle=\left\langle p\right\rangle=0,\qquad G\left(\phi_{i}\right)=0.
\end{equation}\\
  If $\zeta<0$ and $2\leq d\leq 3$, we have a $\mathbb{Z}_{d}$-orbifolded LG model
\begin{equation}\label{LGAGLSM}
  \widehat{W}:=\sum^{N+1}_{i=1}\Phi^{d}_{i},\qquad \left\langle \sigma\right\rangle=0,\qquad \left\langle p\right\rangle=\sqrt{-\frac{\zeta}{d}}.
\end{equation}\\

\item   The topological vacua are given by solving $F(\sigma)=\zeta+k\left(\sigma-|\sigma|\right)=0$
\begin{equation}\label{tvgs}
  \langle\sigma\rangle=-\frac{\zeta}{2k}<0.
\end{equation}
\end{itemize}

So if $k>0$, the topological vacua are in the geometric phase, while, if $k<0$, the topological vacua are at the LG point. The low energy effective theory is a $U(1)_{d^{2}-N-1}$ Chern-Simons theory.  Now, we compute the Witten index for several examples.

\subsubsection{GLSM for $\mathbb{P}^{N}[2]$}
In section \ref{SEW}, we discussed how Witten's analysis can be extended in the phases of the GLSM for $\mathbb{P}^{N}[2]$ on the spacetime $\RR^2\times S^{1}$. Now, we analyze the Witten index of the phases of the 3d GLSM for $\mathbb{P}^{N}[2]$ before compactification. Let us start with the topological vacua first. We assume $N>3$, and because $d=2$, the effective $U(1)_{3-N}$ Chern-Simons theory is at the LG point. We find the Witten index is $N-3$ by counting the dimension of the Verlinde algebra. As the dimension of the K-group of $\mathbb{P}^{N}[2]$ is $N$ if $N$ is even and $N+1$ if $N$ is odd, this leads to the first surprise: the Verlinde formula of the $U(1)_{3-N}$ Chern-Simons theory can not match the dimension of the K-group of the target for any $N$.

In order for the Witten index to be invariant under phase transition, the LG model must have the number of vacua equal to 3 for even $N$ and 4 for odd $N$. This suggests that the Witten index of the three-dimensional orbifolded Landau-Ginzburg model is different from its two-dimensional counterpart. Now, we give an intuitive answer of the Witten index of the 3d $\mathbb{Z}_{2}$ LG orbifolds in this section, and put the formal derivation in appendix \ref{TWLGO}. As in the 2d $\mathbb{Z}_{2}$ LG orbifold, we have ground states:
\begin{equation}\label{3dLG}
   |\Omega\rangle,\qquad |{\rm TS}\rangle_{i}\quad {\rm for}\quad i=1,\ldots,3.
\end{equation}
The untwisted sector $|\Omega\rangle$, which can again be represented as $ d\phi_{1}\wedge\cdots\wedge d\phi_{N+1}$, will survive if $N$ is odd. Now, we have one more spatial coordinate, so there are overall three twisted sectors: the anti-periodic boundary condition of each spatial coordinate gives two twisted sectors plus one twisted sector from the anti-periodic boundary conditions on both coordinates. So, indeed, the 3d $\mathbb{Z}_{2}$-orbifolded LG model has three vacua for even $N$ and four vacua for odd $N$. This means that the Witten index is well-defined in the 3d gauge theory as expected \cite{Intriligator:2013lca}.

It was known that theories related by dimensional reduction generally do not have the same index \cite{Seiberg:1996nz,Aharony:1997bx}. For example, 4d ${\cal N}=1$ $SU(2)$ pure gauge theory has two vacua, whereas 3d ${\cal N}=2$ SYM has no SUSY vacua. This can be understood from the KK reduction of 4d ${\cal N}=1$ gauge theory. The vacua can be represented as critical loci of the superpotential
 \begin{equation}\label{4T3}
   W= \frac{1}{Y}+\eta Y,
 \end{equation}
where $\eta\mapsto 0$, if the radius $R$ of the compactification circle of dimensional reduction goes to zero. One can find the two vacua
\begin{equation}\label{}\nonumber
  \langle Y\rangle=\pm\frac{1}{\sqrt{\eta}}
\end{equation}
both move out to infinity of the field space $Y$.

However, the mismatch of the Witten index between the 3d $\mathbb{Z}_{2}$-orbifolded LG model and its 2d counterpart is different from the above case. The extra two vacua do not depend on the size of $S^{1}$ and can not be expressed in terms of field variables of the LG model. Thus we do not have a similar explanation as in (\ref{4T3}).

Recall in section \ref{SEW}, we put a 3d gauge theory on $\RR^2\times S^{1}$ first, and notice phases of the gauge theory are essentially the same as those of its 2d version. However, in this section, we find that phase structures of a
three-dimensional GLSM on flat $\RR^3$ are, in general, different from its two-dimensional counterpart. So we come to the first observation:
\begin{center}
\textit{ The KK reductions of phases of a 3d GLSM are often different from phases of the 2d GLSM for the same target.}
\end{center}

One consistency check is to study how the vacua are rearranged in taking the decompactified limit of a GLSM on $\RR^2\times S^{1}$. Let us recall the vacuum equation (\ref{VEAB2P}),
\begin{equation}\nonumber
   \left(1-x\right)^{N-1}=q\left(1+x\right)^{2},
 \end{equation}
and solutions of the above equation are the topological vacua of the 3d gauge theory on $\RR^2\times S^{1}$ for $\mathbb{P}^{N}[2]$. Note that
\begin{equation}\label{}\nonumber
  q=\left(\mu R\right)^{N-1}\exp\left(-2\pi R\zeta\right)
\end{equation}
goes to infinity, when $R\rightarrow \infty$ for a negative $\zeta$. In this limit, the solutions of equation (\ref{VEAB2P}) split into two types:
\begin{eqnarray}\label{}\nonumber
  \langle\sigma\rangle&\simeq&\frac{\zeta}{N-3}-\frac{N-1}{N-3}\frac{\ln\left(\mu R\right)}{2\pi R}+\frac{i}{2R}-\frac{il}{R\left(N-3\right)},\quad {\rm for}\quad l=0,\ldots,N-4,\\\nonumber
  \langle\sigma\rangle&\simeq&-\frac{i}{2R}\pm\frac{\sqrt{2^{N-3}}}{\pi}\mu^{-\frac{N-1}{2}}R^{-\frac{N+1}{2}}\exp\left(\pi R\zeta\right).
\end{eqnarray}
There will be $N-3$ degenerate vacua at $\langle\sigma\rangle=\frac{\zeta}{N-3}$ in the decompactification limit, which are exactly the topological vacua of the 3d GLSM defined in equation (\ref{tvgs}). On the other hand, two vacua move out to $\sigma=0$ when $R\rightarrow \infty$, and this agrees with the difference between the Witten index of the 3d $\mathbb{Z}_{2}$-orbifolded LG model and the 2d $\mathbb{Z}_{2}$ LG orbifold.

\subsubsection{GLSM for $\mathbb{P}^{2}[3]$}
Our last abelian example in this section is a GLSM for the elliptic curve $\mathbb{P}^{2}[3]$. The Chern-Simons level we choose for this case is $k=3$, then we find the topological vacua are, at $\zeta>0$, in the geometric phase:
\begin{equation}\label{}\nonumber
  \langle\sigma\rangle=-\frac{\zeta}{6}<0.
\end{equation}
It represent the vacua of a $U(1)_{6}$ Chern-Simons theory, which has six ground states. The LG point of this GLSM is a $\mathbb{Z}_{3}$-orbifolded LG model with superpotential
\begin{equation}\label{3declg}
  W=\Phi^{3}_{1}+\Phi^{3}_{2}+\Phi^{3}_{3}.
\end{equation}

Because the target geometry $\mathbb{P}^{2}[3]$ has vanishing Witten index, this suggests that the LG orbifold defined in (\ref{3declg}) has the Witten index 6. Indeed, it agrees with the computation in appendix \ref{TWLGO}. This 3d example is surprising because there should be no ``topological vacua" in the 2d GLSM for a Calabi-Yau manifold. So one may suspect that the existence of the extra topological vacua in 3d is simply due to our special choice of the Chern-Simons level. However, this is not the case. If one chooses a different level, then the geometric phase and the LG point will both have topological vacua, and the difference between them is still 6.

This new phenomenon also appears in the case of more general Calabi-Yau targets, so it may have important applications to the K-theoretic Gromov-Witten theory, which we leave to future work.

\section{Decompositions after KK reductions}\label{DKK}
In the previous section, we observed that 2d LG orbifolds cannot be simply regarded as the KK reductions of their 3d counterparts. In particular, we computed the Witten indices of 2d/3d LG orbifolds and found they are different. Furthermore, the ground states of 3d LG orbifolds do not have an explicit dependence on the radius, so it is unclear how these vacua of the 3d theory evolve when the radius goes to zero. In this section, we discuss another phenomenon in which one can see the detailed connection between 3d and 2d in the KK reduction. Let $G$ be a finite abelian group. We observe that
\begin{center}
    \textit{The circle reduction of a 3d gauge theory for the G-gerbe over projective space with a suitable Chern-Simons level will decompose into} $|G|$  \textit{of 2d GLSMs for the same target}.
\end{center}
The notation of $G$-gerbe over projective space is a special orbifold in math, and we mainly focus on the case of $G=\mathbb{Z}_{l}$ in this paper. The two-dimensional GLSM for a gerbe was first studied in \cite{Pantev:2005zs,Hellerman:2006zs}, with the upshot being that the gerbe structure suggests that the matter fields in gauge theory have non-minimal charges. For example, for $\mathbb{Z}_{l}$-gerbe over a projective space, the matter field in the gauge theory have charges all $l$.
This 2d study can be easily extended into a three-dimensional gauge theory, and one can study the dynamics as well as the KK reduction of the 3d gauge theory for this target in the context of the ``generalized global symmetries" \cite{Gaiotto:2014kfa} which act on non-local operators like Wilson/'t Hooft loops \cite{Witten:1999ds}.\footnote{For example, a 3d abelian gauge theory with matter fields with charges all divisible by $l$ will have a $\ZZ_l$ 1-form symmetry. Such 1-form symmetry of the system on $\RR^2\times S^1$ predicts the existence of different sectors of the effective 2d theory. This 1-form symmetry in the 3d theory can be identified with the dual symmetry that arises after gauging the $\ZZ_l$ symmetry of the ``unorbifolded" theory.} However, in this paper, we will verify this statement by studying the twisted effective superpotential reviewed in section \ref{GLSM32}.

 Finally, we want to give several comments for a general orbifold. If the orbifold, for example the weighted project space $W\mathbb{P}_{[1,2]}$, can still be regarded as the Higgs vacuum of a gauge theory, we expect that there is a similar decomposition statement in the KK reduction. However, the decomposed 2d theories are usually different to each other. We will come back to this point at the end of this section. If the orbifold cannot be embedded into a gauge theory, just like the Landau-Ginzburg orbifolds, we do not know whether we have a dynamical decomposition in the circle reduction. More specifically, although the vacua as a vector space could be ``decomposed," it is not clear to us whether its Hermitian structure has the same decomposition in the KK reduction. 

\subsection{3d GLSM for $W\mathbb{P}_{[l,\cdots,l]}$}
Without loss of generality, we start with the 3d GLSM for $W\mathbb{P}_{[l,\cdots,l]}$. It is a $U(1)$ gauge theory with $N+1$ charge-$l$ matter fields, and the Chern-Simons level is chosen to be $-\frac{l^{2}\left(N+1\right)}{2}$ such that there are no topological vacua in the geometric phase. The common factor $l$ in the gauge charges indicates that it is a $\mathbb{Z}_{l}$-orbifolded theory. This can be seen from that, in the geometric phase, the gauge group $U(1)$ is not fully Higgsed by the vacuum expectation values of matter fields and left with a $\mathbb{Z}_{l}$ gauge theory. 
In this section, we study the theory at the LG point where one can integrate out all matter fields. If we put the theory on spacetime $\RR^2\times S^{1}$, we then reduce the Chern-Simons-matter theory to an effective theory described by the twisted effective superpotential,
\begin{equation}\label{GLTWS}
  \widetilde{W}^{3\dd}_{\eff}\left(\Sigma, t\right)\cdot 2\pi R=\frac{1}{2}\left(k+\frac{l^{2}\left(N+1\right)}{2}\right)\left(\ln x\right)^{2}+\left(\ln q\right)\left(\ln x\right)+\left(N+1\right){\rm Li}_{2}\left(x^{l} \right).
\end{equation}
The vacuum equation is then
\begin{equation}\label{vel}
  (1-x^{l})^{l(N+1)}=q_{3\dd}.
\end{equation}
This equation can be solved by
\begin{eqnarray}\label{vel2}
  \langle \Sigma\rangle&=&-\frac{1}{2\pi Rl}\ln\left(1-Rq^{\frac{1}{l(N+1)}}_{2d}e^{\frac{2i\pi m}{l(N+1)}}\right)-\frac{i}{Rl}n, \\\nonumber
  && {\rm for}\ m=0,\ldots,l(N+1)-1,\ {\rm and}\ n=0,\ldots,l-1.
\end{eqnarray}
In the above equation, we have used the relation between the 3d parameter and its 2d version
\begin{equation}\label{}\nonumber
  q_{3\dd}=R^{l(N+1)}q_{2\dd}.
\end{equation}
 One can then read that the Witten index is $l^{2}(N+1)$. Now, we fix $q_{2\dd}$ and investigate how these vacua distribute in the 2d limit where $R$ goes to zero:
\begin{equation}\label{}\nonumber
   \langle \Sigma\rangle\simeq\frac{1}{2\pi l}q^{\frac{1}{l(N+1)}}_{2\dd}e^{\frac{2i\pi m}{l(N+1)}}-\frac{i}{Rl}n, \quad {\rm for}\ m=0,\ldots,l(N+1)-1,\ {\rm{and}}\ n=0,\ldots,l-1.
\end{equation}
One observes that the case of $n=0$ gives vacuum solutions of 2d GLSM for $W\mathbb{P}_{[l,\cdots,l]}$:
 \begin{equation}\nonumber
   \langle \Sigma\rangle=\frac{1}{2\pi l}q^{\frac{1}{l(N+1)}}_{2\dd}e^{\frac{2i\pi m}{l(N+1)}}, \quad {\rm for}\ m=0,\ldots,l(N+1)-1.
 \end{equation}
The above vacua can be obtained from the 2d twisted effective superpotential of GLSM for $W\mathbb{P}_{[l,\cdots,l]}$ as well
\begin{equation}\nonumber
  \widetilde{W}^{2\dd}_{\eff}\left(\Sigma,t\right)=\left(\ln q_{2\dd}\right)\left(\ln x\right)-l\left(N+1\right)\log\Sigma.
\end{equation}
For two different $n$'s, one may naively observe that they are infinitely far away on the $\Sigma$ field space, which means they will decouple each other in the 2d limit. To justify this physical expectation, we will estimate the mass of the soliton spectrum between two different $n$'s. Following \cite{Hori:2000ck}, we can calculate
\begin{equation}\label{}\nonumber
  M_{n_{1}n_{2}}\simeq \left|\widetilde{W}^{3\dd}_{\eff}\left(\Sigma_{n_{1}}\right)-\widetilde{W}^{3\dd}_{\eff}\left(\Sigma_{n_{2}}\right)\right|\simeq \left|\frac{i\left(n_{2}-n_{1}\right)}{Rl}+{\cal O}(R^{0})\right|,
\end{equation}
 where we have implicitly used the K\"{a}hler metric of $\Sigma$ field space investigated in \cite[section 4.2]{Aganagic:2001uw}. From above, one finds that the soliton spectrum between two vacua with different $n_{i}$'s will become infinitely heavy, and their dynamics are frozen. So, indeed, the 3d GLSM for $W\mathbb{P}_{[l,\cdots,l]}$ decomposes into $l$ identical 2d GLSMs for $W\mathbb{P}_{[l,\cdots,l]}$ in the 2d limit. Different sectors are labeled by different $n$'s, which can be regarded as the charge under the $\mathbb{Z}_{l}$ 1-form global symmetry.

 We conclude this section by mentioning 
 some related discussions on this phenomenon in the literature. For example in \cite{Aharony:2017adm}, they found that a 3d ${\cal N}=2$ gauge theory would decompose into a direct sum of different 2d theories; see also \cite{Hwang:2018riu}. Our observation provides another possibility that the decomposed 2d theories are identical and their target space is the same as the one in the 3d theory.

 Notice that the decomposition also applies when the charges are not identical but share the same common divisor. For example, the 3d $\mathcal{N}=2$ gauge theory for $W\PP_{[2,4]}$ decomposes into two identical copies of 2d gauge theories in the 2d limit. Each copy of the 2d gauge theories further decomposes into two sectors with one sector equal to the 2d GLSM for $W\PP_{[2,4]}$.

\section{Conclusions and future directions}
In this paper, we studied phases of 3d ${\cal N}=2$ Chern-Simons-matter theory with a renormalizable superpotential. Although our original motivation was to understand the correspondence between the Verlinde formula and the quantum K-theory of a geometric target space in the three-dimensional gauge theory, we found several striking physical aspects which were not studied carefully in the literature before. The first observation is that the Landau-Ginzburg orbifolds obtained from 3d ${\cal N}=2$ GLSMs with a superpotential on $\RR^2\times S^{1}$ is different from phases of the 3d flat spacetime theory. We computed the Witten index of 3d/2d LG orbifolds to demonstrate this observation. Furthermore, it is unclear, at the current stage, about the relation between the 3d LG orbifold and its 2d version because the extra twisted sectors of 3d orbifolds do not depend on the radius of $S^{1}$ in the KK reduction. However, for a gauge theory for a geometric orbifold, the connection between the 3d gauge theory and its 2d version is clear. We provided an explicit connection by studying a concrete example. We computed the twisted effective superpotential of 3d GLSM for $W\mathbb{P}_{[l,\cdots,l]}$ and observed that the theory will decompose into $l$ copies of 2d GLSMs for the same target in the vanishing limit of the circle.

In the study of phases of the 3d gauge theory in this paper, we have turned off real masses of matter fields for having a non-trivial bare superpotential in the action. However, the real mass is an important ingredient in understanding the 3d ${\cal N}=2$ mirror symmetry \cite{Aganagic:2001uw,Intriligator:2013lca}. So it is natural to understand whether our gauge theory could have a 3d mirror dual, and how it is related to the mirror symmetry in the 2d case \cite{Hori:2000kt,Aganagic:2001uw,Gu:2018fpm}. One implication of our observation is that Landau-Ginzburg models which are mirror in 2d need not be mirror in 3d.\footnote{For example, the 2d orbifolded LG model with the superpotential $W=\Phi^{n}/\mathbb{Z}_{n}$ is mirror to the same superpotential without that orbifold.} So it is interesting to understand what modification is needed in 3d to preserve the duality. Another interesting question is to ``categorify'' the Verlinde/Grassmannian correspondence to an equivalence between tensor categories using the full 3d theory as opposed to the effective theory on $\RR^2\times S^1$  (cf. \cite[sec.~2]{Gukov:2016gkn}). We leave all these to future work.

Finally, we want to comment that the observations we made in this paper suggest that lifting a 2d theory to a 3d theory has to be done with care and caution,\footnote{There is a recent work \cite{Jockers:2018sfl} on GLSMs on $\RR^2\times S^{1}$ for more general targets such as the quintic. Its relation to 3d gauge theories on flat spacetime is unclear to us at the moment.} as the 3d theory often contains additional information that got lost under compactification.
Thus, it would be useful to give criteria for when a lifting process is ``canonical.'' We hope to further investigate this in the future.

\section*{Acknowledgement}
We would like to thank A. Kapustin, Y. Ruan and N. Seiberg for conversations. We especially thank S. Gukov and E. Sharpe for carefully reading the draft and providing valuable feedback.

\appendix
\section{The Witten index of 3d ${\cal N}=2$ Landau-Ginzburg orbifolds}\label{TWLGO}
In this appendix, we provide a formal method of calculating the Witten index of 3d ${\cal N}=2$ Landau-Ginzburg orbifolds. The Witten index of a  three-dimensional theory is defined by computing the three-torus partition function
\begin{equation}\label{3dWi}
  Z_{T^{3}}={\rm Tr}_{T^{3}}(-)^{F}.
\end{equation}
For a non-orbifolded theory described by (\ref{3dWi}), all fields are periodic. However, if it is a $G$-orbifolded theory, for a finite group $G$, there could be ground states from twisted boundary conditions which are so-called twisted sectors. To take into account their contributions, we need to sum over all non-trivial principal $G$-bundles over the 3-torus where the gauge theory lives on. This can be readily calculated by \cite{Dijkgraaf:1989pz}
\begin{equation}\label{}
  {\rm Hom}\left(\pi_{1}\left(T^{3}\right), G\right)/G.
\end{equation}
Then elements of the above set are commuting triples,
\begin{equation}\label{}\nonumber
  \left\{\left(g,h,k\right)\in G^{3}:[g,h]=[h,k]=[k,g]=1_{G}\right\}/G.
\end{equation}
So the orbifold partition function can be evaluated to give
\begin{equation}\label{3opf}
  Z_{T^{3}}=\frac{1}{|G|}\sum_{g,h,k\in G}Z^{{\rm Orb}}_{T^{3}}\left(g,h,k\right).
\end{equation}
Recall that the two-torus partition function is invariant under the $SL(2,\mathbb{Z})$ modular transformation of the two-torus with complex structure $\tau$:
\begin{equation}\label{2MS}
  \tau\mapsto\frac{a\tau+b}{c\tau+d}, \quad ad-bc=1,\ {\rm and}\ a,b,c,d\in\mathbb{Z}.
\end{equation}
To be more precise, we have
\begin{equation}\label{2pmi}
  Z^{{\rm Orb}}_{T^{2}}\left(h,k\right)=Z^{{\rm Orb}}_{T^{2}}\left(h^{a}k^{b},h^{c}k^{d}\right).
\end{equation}
Since our goal here is to compute $\mathbb{Z}_{N}$-orbifolded LG models, we can always have a transformation such that $h^{c}k^{d}=1_{G}$.
If this transformation is the special transformation $S$ of two-torus, then $h$ must be the identity group element $1_{G}$. For a more general group element $h\neq 1_{G}$, this suggests $d\neq 0$, and we have
\begin{equation}\label{2pmi}
  Z^{{\rm Orb}}_{T^{2}}\left(h,k\right)=Z^{{\rm Orb}}_{T^{2}}\left(w,1_{G}\right), \qquad w\neq1_{G}.
\end{equation}
Notice that if $k=1_{G}$, the partition function $Z^{{\rm Orb}}_{T^3}\left(g,h,1_{G}\right)$ is essentially a two-torus partition function $Z^{{\rm Orb}}_{T^2}\left(g,h\right)$ as the Witten index only depends on constant modes of field spaces. Furthermore, because the mapping class group of two-torus $SL(2,\ZZ)$ can be regarded as a $2\times2$ block inside the 3-torus mapping class group $SL(3,\mathbb{Z})$, we can then use the constraint (\ref{2pmi}) to express the partition function (\ref{3opf}) as a summation over orbifolded two-torus partition functions,
\begin{equation}\label{3tp2tp}
  Z_{T^{3}}=\frac{1}{|G|}\left\{\sum_{g,h\in G}Z^{{\rm Orb}}_{T^{3}}\left(g,h,1_{G}\right)+\sum_{\{g,k\neq 1_{G}\}\in G}Z_{T^{3}}^{{\rm Orb}}\left(g,k,1_{G}\right)+\sum_{\{g,h\neq 1_{G},k\neq 1_{G}\}\in G}Z_{T^{3}}^{{\rm Orb}}\left(g,w,1_{G}\right)\right\}.
\end{equation}
Thus the problem has been reduced to a computation of two-torus orbifolded partition functions. Although the above expression is enough for the $\mathbb{Z}_{N}$ orbifold, we comment that for a general orbifold, one may need the entire mapping class group of 3-torus $SL(3,\mathbb{Z})$ to get relations among twisted sectors, which we leave to the interested reader.

For 2d LG orbifolds, the partition functions have been studied in the literature \cite{Vafa:1989xc,Kawai:1993jk}, and we briefly summarize the main results we will use in \cite{Vafa:1989xc}. The superpotential $W$ in the LG model is a weighted homogeneous polynomial of $N$ chiral superfields $\Phi_{1},\cdots,\Phi_{N}$ with R-charges $q_{1},\cdots,q_{N}$,
\begin{equation}\label{}
  \lambda^{2}W\left(\Phi_{1},\cdots,\Phi_{N}\right)=W\left(\lambda^{q_{1}}\Phi_{1},\cdots,\lambda^{q_{N}}\Phi_{N}\right).
\end{equation}
We require that $W$ has an isolated critical point at the origin and $q_{i}$'s are strictly positive rational numbers such that all $q_{i}\leq 1$. The orbifolded LG partition function is given by \cite{Vafa:1989xc}
\begin{equation}\label{2dolpf}
  Z^{{\rm Orb }}_{T^{2}}=\frac{1}{|G|}\sum^{|G|-1}_{m,n=0}\left(-1\right)^{\left(\widehat{c}+N\right)\left(m+n+mn\right)}\prod_{\frac{mq_{i}}{2},\frac{nq_{i}}{2}\in\mathbb{Z}}\left(1-\frac{2}{q_{i}}\right),
\end{equation}
where $\widehat{c}$ is one third of the SCFT central charge of the LG model: $\widehat{c}=c/3=\sum^{N}_{i=1}\left(1-q_{i}\right)$.

Now we compute the Witten index for several 3d Landau-Ginzburg orbifolds.

\begin{itemize}
    \item   A $\mathbb{Z}_{2}$ LG orbifold with the superpotential $W=\sum^{N+1}_{i=1}\Phi^{2}_{i}$.
One can calculate the two-dimensional Witten index
\begin{equation}\label{}\nonumber
  Z^{{\rm \mathbb{Z}_{2} }}_{T^{2}}=\frac{1}{2}\left(\left(-1\right)^{N+1}+3\right).
\end{equation}
Thus, the Witten index is 2 if $N$ is odd, and 1 if $N$ is even. The three-dimensional Witten index can also be computed
\begin{equation}\label{}\nonumber
  Z^{{\rm \mathbb{Z}_{2} }}_{T^{3}}=\frac{1}{2}\left(\left(-1\right)^{N+1}+7\right).
\end{equation}
So we have 4 ground states for odd $N$ and 3 vacua for even $N$.

\item A $\mathbb{Z}_{3}$ LG orbifold with the superpotential $W=\sum^{3}_{i=1}\Phi^{3}_{i}$.
The two-dimensional Witten index of the untwisted sector is
\begin{equation}\label{}\nonumber
  Z_{T^{2}}(0,0)=-8,
\end{equation}
and each of the eight twisted sectors contributes 1, so the 2d Witten index vanishes as expected.
The three-dimensional Witten index is
\begin{equation}\label{}\nonumber
  Z^{{\rm \mathbb{Z}_{3} }}_{T^{3}}=\frac{1}{3}\left(\sum^{2}_{m,n=0}Z^{{\rm \mathbb{Z}_{3} }}_{T^{3}}(m,n,0)+\sum^{2}_{m,n=0}Z^{{\rm \mathbb{Z}_{3} }}_{T^{3}}(m,n,1)+\sum^{2}_{m,n=0}Z^{{\rm \mathbb{Z}_{3} }}_{T^{3}}(m,n,2)\right)=\frac{1}{3}\left(0+9+9\right)=6.
\end{equation}

\item Our last example is a $\mathbb{Z}_{4}$ LG orbifold with the superpotential $W=\sum^{3}_{i=1}\Phi^{4}_{i}$. This LG model can have a well-defined UV Lagrangian, although it could not be embedded into a flat three-dimensional GLSM. The two/three-dimensional Witten index can be readily computed following the setup in this appendix and they are 24 and 36 respectively.

\end{itemize}

Finally, it can be seen that three-dimensional orbifolds usually have more ground states than their two-dimensional version. All these extra ground states in 3d are from the twisted sectors and it would be interesting to investigate the quantum numbers of those states such as the R-charge \cite{Vafa:1989xc}. One may naturally expect that the extra ground states may share the same quantum numbers as homogeneous polynomials of $\sigma$'s field in the 3d GLSM, and we leave this investigation to the interested reader.

\end{document}